\documentclass[12pt]{article}
\usepackage{graphicx}
\usepackage{ulem}
\usepackage{cite}

\usepackage{amssymb}

\usepackage{color}

\usepackage{amsmath}
\usepackage[makeroom]{cancel}
\usepackage{amsfonts}

\newcommand{\beq}{\begin{equation}}
\newcommand{\eeq}{\end{equation}}

\begin{document}

\begin{titlepage}

\vskip 2cm
\begin{center}
{\Large\bf  Non-commutativity and non-inertial effects on the Dirac oscillator in a cosmic string space-time
\footnote{{\tt rodrigo.cuzinatto@unifal-mg.edu.br}, {\tt mdemonti@ualberta.ca}, {\tt pompeia@ita.br}}}
 \vskip 10pt
{\bf
R. R. Cuzinatto$^a$, M. de Montigny$^{b}$, P. J. Pompeia$^{c}$ }
\vskip 5pt
{\sl $^a$Instituto de Ci\^encia e Tecnologia, Universidade Federal de Alfenas\\
Rodovia Jos\'e Aur\'elio Vilela, 11999, Cidade Universit\'aria\\
CEP 37715-400 Po\c cos de Caldas, Minas Gerais, Brazil}
\vskip 2pt
{\sl $^b$Facult\'e Saint-Jean, University of Alberta \\
 8406 91 Street NW\\
 Edmonton, Alberta, Canada T6C 4G9, Canada}
 \vskip 2pt
 {\sl $^c$Departamento de F\'\i sica, Instituto Tecnol\'ogico de Aeron\'autica\\
Pra\c ca Mal. Eduardo Gomes 50\\
CEP 12228-900 S\~ao Jos\'e dos Campos, S\~ao Paulo, Brazil}
\end{center}

\begin{abstract}
We examine the non-inertial effects of a rotating frame on a Dirac oscillator in a cosmic string space-time with non-commutative geometry in phase space.  We observe that the approximate bound-state solutions are related to the biconfluent Heun polynomials. The related energies cannot be obtained in a closed form for all the bound states. We find the energy of the fundamental state analytically by taking into account the hard-wall confining condition. We describe how the ground-state energy scales with the new non-commutative term as well as with the other physical parameters of the system.
\end{abstract}

\bigskip



\end{titlepage}

\newpage

\setcounter{footnote}{0} \setcounter{page}{1} \setcounter{section}{0} %
\setcounter{subsection}{0} \setcounter{subsubsection}{0}

\newcommand{\anu}{\overline\nu}




\section{Introduction{\label{introduction}}}

The main purpose of this paper is to analyze the effects of non-commutativity on the dynamics of a Dirac oscillator in a rotating cosmic string space-time. Our contribution thus adds a novel feature to the work by Bakke \cite{Bakke} which considered the same system but in the standard context of a commutative space-time.

As proposed recently in Ref.  \cite{Brandenberger2013}, observational cosmology is so thriving that its data might help understand particle physics beyond the Standard Model and the reach of present or future particle accelerators, thus `probing particle physics from top down'. An example of such cosmological objects is the concept of cosmic strings, which may occur in some quantum field theories in the form of linear defects in stable field configurations \cite{5,VilenkinShellard,Hindmarsh}.  Although the Standard Model itself does not admit string configurations stable in the vacuum, a quantum field theory beyond the Standard Model which allows cosmic string solutions might have led, according to the Kibble mechanism, to the formation of a network of strings during the cooling process in the early universe, that might continue to the present time \cite{6}. 

The cosmic strings are spatial lines of trapped energy density, analogous to vortex lines in superfluids and superconductors or line defects in crystals \cite{Brandenberger2014}. The cosmological signatures of the gravitational effects caused by this energy density are expected to be quite distinctive, and it is therefore important to understand the interaction between relativistic fields and these gravitational fields. The influence of defects was investigated in many papers dealing,  for example, with the scattering of particles \cite{1,Deser,18}, single particles in various potentials within space-times with topological defects \cite{4}, or the harmonic oscillator interacting with topological defects \cite{2,3}.  The term `Dirac oscillator' was introduced in 1989 \cite{Moshinsky}, although a similar equation had been introduced in 1967 \cite{Ito} and solved four years later \cite{Cook}. It is an exactly solvable model introduced in the context of many-particle models in relativistic quantum mechanics. The associated equation, defined by the Dirac equation with an imaginary effective linear vector potential added to the momentum operator, followed the rationale that since the oscillator potential is quadratic in the coordinates, the relativistic description of spin-half particles should involve an equation linear in both coordinates and momenta. In addition to being a rare solvable relativistic model, its non-relativistic limit is the Schr\"odinger harmonic oscillator with in addition a strong spin-orbit coupling. A recent comment with early references can be found in Ref. \cite{Quesne}. Recent contributions which describe the influence of non-inertial effects on the Dirac oscillator in the cosmic string space-time background are in Refs. \cite{Bakke,Bakke1,Bakke2} and the references therein. The Dirac oscillator in a topological defect space-time was examined in Refs. \cite{Bakke, Bakke1, Bakke2,Villalba,Mandal1,Moraes,BakFur,Mandal2,Panella,Nath,Mota,Deng,Laba,Salazar,Oliveira}.  The bound states of electrons and holes to disclinations were investigated in Refs. \cite{N20,N21}, as well as the calculations of Landau levels in the presence of topological defects \cite{8,N23,N24}.

To our knowledge, the first major paper which exploited the idea that configuration-space coordinates do not commute is Ref. \cite{discrete}.  It may be interesting to note that the idea first came to Heisenberg as a prospective remedy for short-distance singularities; he mentioned his idea to Peierls, who relayed it to Pauli, who relayed it to Oppenheimer, who asked his student H. S. Snyder to explore this idea \cite{Jackiw}. About twenty years ago, there was a revived interest in non-commutative quantum mechanics in the study of the low-energy effective theory of D-branes in the background of a Neveu-Schwarz B-field in a non-commutative space  \cite{NC,NC2,NC3,NCQM}. Recent investigations of non-commutative geometry pertained to the quantum Hall effect  \cite{Hall,Hall2,Hall3,Hall4}, the Landau problem \cite{Landau,Landau2,Landau3,Landau4,Landau5},  planar quantum systems with central potentials \cite{planar,planar2}, geometric phases \cite{Passos, Ribeiro}, the Dirac oscillator \cite{Mandal2,Panella,Cai,Zarrinkamar,Hou} and the relativistic Duffin-Kemmer-Petiau oscillator \cite{Yang,Guo,Falek}.  Finally, let us mention that the scattering of non-commutative vortices was investigated in Ref. \cite{Joseph} in order to describe the interaction of cosmic strings. On the other hand, our paper deals with the behaviour of a Dirac oscillator in the presence of a single cosmic string and non-commutativity. In particular, we will be interested in analyzing the effects of the non-commutativity in the phase space, specially concerning the momenta commutation relations. The interesting point of this analysis is that the non-commutativity of the momentum coordinates emulates the presence of a background magnetic field \cite{Landau5,NCFT}.

The wave function of the Dirac field considered here can be approximated in terms of Heun polynomials. The Heun function is considered as a generalization of the ``hypergeometric, Lam\'e, Mathieu, spheroidal wave and many other known special functions" \cite{HF2I}. It has had an increasing interest in physical applications in the last decades. In fact, there is a plethora of physical systems in which the Heun functions appear like in the analysis of black holes, braneworld, quantum field theory \cite{HFApp1,HFApp2,HFApp3,HFApp4,HFApp5} and many others (see Ref. \cite{HFApp6} for a long list of applications). In this sense, our analysis reinforces the importance of the Heun functions in physics.  In particular, of special interest here are the Heun polynomials, which are related to the energy eigenfunctions and the quantized energy of the system.

In this paper, we investigate the effects of non-commutativity in a cosmic string space-time equipped with a rotating frame in cylindrical coordinates by writing the Dirac equation in this curved space-time. As mentioned, e.g. in Ref. \cite{Bakke} and references therein, the geometry of the space-time can play the role of a hard-wall confining potential via non-inertial effects. We consider non-commutativity  in the momentum components only; this modifies the non-minimal oscillator prescription by adding a new term related to the Bopp shift. We study the bound-state solutions and find the corresponding energies. In Section \ref{SecCosmicString}, we obtain Eq. (\ref{Equation1}) for the Dirac oscillator in a rotating cosmic string space-time with non-commutative momenta.  In Section \ref{Solutions}, we investigate the solutions by restricting the non-commutativity parameters to be parallel to the cosmic string. We use the Frobenius method to determine an approximate solution of the wave function in terms of the Heun polynomials and, most importantly, the energy eigenvalues, which depend on the string parameters, rotation parameter and non-commutativity parameter. Due to the constraint for the termination of the Frobenius series, we are not able to get a closed expression for the energy levels which is valid for all the values of the principal quantum number; instead, each energy level has to be obtained by solving the quantization condition together with the constraint equation order-by-order. Here, we exemplify the procedure for the ground energy level of our system. As discussed in Section \ref{Solutions}, the approximate nature of the solution is also due to the presence of the harmonic oscillator subject to a hard-wall confining potential. We describe the validity of our approximation in statistical terms, that is, we express the statistical error from the hard-wall condition in terms of the parameters which describe our physical system. We briefly discuss the low angular velocity limit of the rotating frame, the large-mass limit and the commutative limit of the model. 




\section{Dirac equation for the oscillator in a non-commutative cosmic string space-time{\label{SecCosmicString}}}

The purpose of this section is to obtain Eq. (\ref{Equation1}) by introducing the Dirac oscillator via a non-minimal coupling within the Dirac equation generalized to curved space-time with a cosmic string metric in a rotating frame, with non-commutative momenta.  Eq. (\ref{Equation1}) is expressed in cylindrical coordinates and involves many parameters: frequency of the rotating frame, a parameter related to the deficit angle, mass of the Dirac field, frequency of the Dirac oscillator, non-commutativity parameters in the momentum space, and the energy. 


In this section, we consider a rotating reference frame, which was studied also in Refs. \cite{Tsai,Cui,Merlin,Vignale,Bakke2010,Mota2014,Castro,Mota2017,Vitoria,BakkeFurtado2009,
BakkeFurtado2010}.
Hereafter, we work with the same cosmic string space-time as in Ref. \cite{Bakke}, with metric signature diag$(-1,+1,+1,+1)$, and described with cylindrical coordinates $(\rho, \varphi, z)$  by
\beq
ds^{2}=-\left(1-\omega^{2}\eta^{2}\rho^{2}\right)dt^{2}+2\omega\eta^{2}\rho^{2}d\varphi dt+d\rho^{2}+\eta^{2}\rho^{2}d\varphi^{2}+dz^{2},
\label{CosmicString}
\eeq
where $\omega$ is the angular frequency of the rotating frame, $\eta=1-4\Lambda$ (with $\Lambda$ the string's linear mass density) runs in the interval $(0,1]$ and is related to the deficit angle $\theta=2\pi\left(1-\eta\right)$, and with units such that $c=1$.  Geometrically, the line element in Eq. (\ref{CosmicString}) describes a Minkowski space-time with a conical singularity \cite{deMello2004}.  The first term in Eq.  (\ref{CosmicString}) implies the natural occurence of two intervals, delimited by $\rho_0\equiv\frac1{\omega\eta}$: $0<\rho<\rho_0$, considered hereafter, and $\rho>\rho_0$, for which the particle lies outside of the light-cone as its velocity is greater than the velocity of light. This involves the condition that the wave function of the Dirac particle must vanish as $\rho$ approaches $\rho_0$. Thus $\omega\eta$ determines two classes of solutions: a finite wall $\rho_0$ with $\omega\eta$ arbitrary but finite and the wall $\rho_0$ at infinity when $\omega\eta\ll 1$ \cite{Bakke}.  


The Dirac equation, with flat Minkowski space-time coordinates $x^a$, for a fermion with mass $m$ is \cite{Greiner}
\beq
i\gamma^{a}\partial_{a}\Psi-m\Psi  =  0.
\label{Dirac}\eeq 
We utilize the standard Dirac $\gamma^a$ matrices in the Minkowski space-time,
\beq
\gamma^0=\left(
\begin{array}{cc}
1 & 0\\
0 & 1\end{array}
\right),\quad
\gamma^j=\left(
\begin{array}{cc}
0 & \sigma^j\\
-\sigma^j & 0\end{array}
\right),\qquad j=1, 2, 3, 
\eeq
where $1$ is the $2\times 2$ unit matrix and 
\beq
\sigma^1=\left(\begin{array}{cc} 0 & 1 \\ 1 & 0\end{array}\right),\quad \sigma^2=\left(\begin{array}{cc} 0 & -i \\ i & 0\end{array}\right),\quad \sigma^3=\left(\begin{array}{cc} 1 & 0 \\ 0 & -1\end{array}\right)\label{Pauli}
\eeq 
are the Pauli matrices.

Its curved space-time version is obtained by means of the tetrad (or vierbein) field $e^a=e^a_{\; \mu}(x)\; dx^\mu$, which can be found from the metric
\[
ds^2=g_{\mu\nu}\; dx^\mu\otimes dx^\nu=\overbrace{\eta_{ab}\; e^a_{\; \mu}(x) e^b_{\; \nu}(x)}^{g_{\mu\nu}}\; dx^\mu\otimes dx^\nu = \eta_{ab}\; e^a\otimes e^b.
\]
The following tetrad components \cite{BakkeFurtado2009,BakkeFurtado2010}, 
\begin{eqnarray}
& e_{\,\mu}^{a}\left(x\right)  =  \left(\begin{array}{cccc}
\sqrt{1-\omega^{2}\eta^{2}\rho^{2}} & 0 & -\frac{\omega\eta^{2}\rho^{2}}{\sqrt{1-\omega^{2}\eta^{2}\rho^{2}}} & 0\\
0 & 1 & 0 & 0\\
0 & 0 & \frac{\eta\rho}{\sqrt{1-\omega^{2}\eta^{2}\rho^{2}}} & 0\\
0 & 0 & 0 & 1
\end{array}\right),\nonumber\\
& e_{\,a}^{\mu}\left(x\right)=\left(\begin{array}{cccc}
\frac{1}{\sqrt{1-\omega^{2}\eta^{2}\rho^{2}}} & 0 & \frac{\omega\eta\rho}{\sqrt{1-\omega^{2}\eta^{2}\rho^{2}}} & 0\\
0 & 1 & 0 & 0\\
0 & 0 & \frac{\sqrt{1-\omega^{2}\eta^{2}\rho^{2}}}{\eta\rho} & 0\\
0 & 0 & 0 & 1
\end{array}\right),\label{Tetrads}
\end{eqnarray}
are consistent with the metric in Eq. (\ref{CosmicString}).

In curved space-time, the gamma matrices of Eq. (\ref{Dirac}) are replaced by
\[
\gamma^\mu=e^a_{\; \mu}\; \gamma^a,
\]
and the derivative $\partial_{\mu}$ is replaced by the covariant derivative,
\beq
\nabla_{\mu}=\partial_{\mu}-\Gamma_{\mu}=e_{\;\mu}^{a}\partial_{a}-\Gamma_{\mu}.
\label{CovariantDerivative}
\eeq
Therefore, the Dirac equation in curved space-time reads
\beq
i\gamma^{\mu}\nabla_{\mu}\Psi-m\Psi=0.
\label{DiracCurved}\eeq

The covariant derivative in Eq. (\ref{CovariantDerivative}) involves the spinorial affine connection  $\Gamma_{\mu}$,
\beq
\Gamma_{\mu}=\frac{i}{2}\omega_{\mu ab}\Sigma^{ab},
\label{SpinorialConnection}\eeq
where $\omega_{\mu ab}$ are the spin connection components
\beq
\omega_{\nu\ d}^{\:b}  =  e_{\:\beta}^{b}\Gamma_{\nu\mu}^{\beta}e_{\:d}^{\mu}-\left(\partial_{\nu}e_{\:\mu}^{b}\right)e_{\:d}^{\mu},
\label{SpinConnection}
\eeq
and $\Sigma^{ab}$ is the commutator
\beq
\Sigma^{ab}=\frac{i}{4}\left[\gamma^{a},\gamma^{b}\right].\label{Sigma-ab}
\eeq
In Eq. (\ref{SpinConnection}), $\Gamma_{\nu\mu}^{\beta}$ denotes the Christoffel symbols which, for the metric in Eq. (\ref{CosmicString}), are
\[
\Gamma_{\nu\mu}^{1}=\left(\begin{array}{cccc}
-\eta^{2}\rho\omega^{2} & 0 & -\eta^{2}\rho\omega & 0\\
0 & 0 & 0 & 0\\
-\eta^{2}\rho\omega & 0 & -\eta^{2}\rho & 0\\
0 & 0 & 0 & 0
\end{array}\right),
\]
\[
\Gamma_{\nu\mu}^{2}=\left(\begin{array}{cccc}
0 & \frac{\omega}{\rho} & 0 & 0\\
\frac{\omega}{\rho} & 0 & \frac{1}{\rho} & 0\\
0 & \frac{1}{\rho} & 0 & 0\\
0 & 0 & 0 & 0
\end{array}\right),
\]
as well as $\Gamma_{\nu\mu}^{0}=0=\Gamma_{\nu\mu}^{3}$.

For the tetrads in Eq. (\ref{Tetrads}) we have
\begin{eqnarray*}
\omega_{0\,b}^{\,a}=\left(\begin{array}{cccc}
0 & -\frac{\omega^{2}\eta^{2}\rho}{\sqrt{1-\omega^{2}\eta^{2}\rho^{2}}} & 0 & 0\\
-\frac{\omega^{2}\eta^{2}\rho}{\sqrt{1-\omega^{2}\eta^{2}\rho^{2}}} & 0 & -\frac{\omega\eta}{\sqrt{1-\omega^{2}\eta^{2}\rho^{2}}} & 0\\
0 & \frac{\omega\eta}{\sqrt{1-\omega^{2}\eta^{2}\rho^{2}}} & 0 & 0\\
0 & 0 & 0 & 0
\end{array}\right) & ,\quad & \omega_{1\,b}^{\,a}=\ensuremath{\left(\begin{array}{cccc}
0 & 0 & \frac{\eta\omega}{1-\eta^{2}\rho^{2}\omega^{2}} & 0\\
0 & 0 & 0 & 0\\
\frac{\eta\omega}{1-\eta^{2}\rho^{2}\omega^{2}} & 0 & 0 & 0\\
0 & 0 & 0 & 0
\end{array}\right),}\\
\omega_{2\,b}^{\,a}=\left(\begin{array}{cccc}
0 & -\frac{\eta^{2}\rho\omega}{\sqrt{1-\eta^{2}\rho^{2}\omega^{2}}} & 0 & 0\\
-\frac{\eta^{2}\rho\omega}{\sqrt{1-\eta^{2}\rho^{2}\omega^{2}}} & 0 & -\frac{\eta}{\sqrt{1-\eta^{2}\rho^{2}\omega^{2}}} & 0\\
0 & \frac{\eta}{\sqrt{1-\eta^{2}\rho^{2}\omega^{2}}} & 0 & 0\\
0 & 0 & 0 & 0
\end{array}\right) & ,\quad & \omega_{3\,b}^{\,a}=\ensuremath{\left(\begin{array}{cccc}
0 & 0 & 0 & 0\\
0 & 0 & 0 & 0\\
0 & 0 & 0 & 0\\
0 & 0 & 0 & 0
\end{array}\right).}
\end{eqnarray*}

From Eqs. (\ref{SpinorialConnection}) and (\ref{Sigma-ab}),  we obtain
\begin{eqnarray*}
\Gamma_{ 0} & = & -\frac{1}{4}\frac{\omega^{2}\eta^{2}\rho}{\sqrt{1-\omega^{2}\eta^{2}\rho^{2}}}\left[\gamma^{0},\gamma^{1}\right]+\frac{1}{4}\frac{\omega\eta}{\sqrt{1-\omega^{2}\eta^{2}\rho^{2}}}\left[\gamma^{1},\gamma^{2}\right],
\end{eqnarray*}

\begin{eqnarray*}
\Gamma_{ 1} & = & \frac{1}{4}\frac{\eta\omega}{\left(1-\eta^{2}\rho^{2}\omega^{2}\right)}\left[\gamma^{0},\gamma^{2}\right],
\end{eqnarray*}

\begin{eqnarray*}
\Gamma_{ 2} & = & \frac{\Gamma_{\left(0\right)}}{\omega}, \qquad \Gamma_{ 3}\ =\ 0.
\end{eqnarray*}


The Dirac oscillator is described by inserting the non-minimal coupling \cite{Bakke,Moshinsky}:
\[
\mathbf{p}\ {\rightarrow}\ \mathbf{p}-im\omega_{0}  \gamma^{0}\mathbf{r},
\]
into the Dirac equation. The frequency of the oscillator is $\omega_0$. In cylindrical coordinates, 
\beq
\mathbf{r}=\rho\hat{e}_{\rho}+z\hat{e}_{z}=\left(\rho,0,z\right).  \label{Cylindrical}
\eeq
Since $p_{i}=-{\rm i}\partial_{i}$ we see that 
\beq
\mathbf{p}\rightarrow-i\left(\nabla+m\omega_{0}  \gamma^{0}\mathbf{r}\right).
\label{Oscillator}\eeq


Non-commutative phase spaces, as the one examined hereafter, can be described by applying a generalized `Bopp shift' Ref. \cite{Melo2013, Curtright} which consists in replacing the coordinates and momenta by the operators\footnote{ Notice we adopt the approach in terms of the Bopp shift. A similar approach is used in Ref. \cite{Landau5}. Other approaches to non-commutativity include the one through the Moyal product. }
\begin{eqnarray}
\hat{r}_{i} & = & r_{i}-\frac{\Theta_{ij}}{2\hbar}p_{j}=r_{i}+\frac{\left(\mathbf{\Theta}\times{\bf p}\right)_{i}}{2\hbar},\nonumber\\
\hat{p}_{i} & = & p_{i}+\frac{\Omega_{ij}}{2\hbar}r_{j}=p_{i}-\frac{\left(\mathbf{\Omega}\times{\bf r}\right)_{i}}{2\hbar},\label{transformation2}
\end{eqnarray}
which implies
\begin{eqnarray*}
\left[\hat{r}_{i},\hat{r}_{j}\right] & = & {\rm i}\Theta_{ij},\\
\left[\hat{p}_{i},\hat{p}_{j}\right] & = & {\rm i}\Omega_{ij},\\
\left[\hat{r}_{i},\hat{p}_{j}\right] & = & {\rm i}\hbar\Delta_{ij},
\end{eqnarray*}
where $\Theta_{i}$ and $\Omega_{i}$ ($i=1,2,3$) are real parameters given by $\Theta_{ij}=\epsilon_{ijk}\Theta_{k}$, $\Omega_{ij}=\epsilon_{ijk}\Omega_{k}$, and
 the parameters $\Delta_{ij}$ are given by 
\[
\Delta_{ij}=\left(1+\frac{\mathbf{\Theta}\cdot\mathbf{\Omega}}{4\hbar^{2}}\right)\delta_{ij}-\frac{\Omega_{i}\Theta_{j}}{4\hbar^{2}}.
\]

As is known in the literature (see for instance Ref. \cite{NCFT}), the non-commutativity of the space leads to a non-locality. In order to emulate the presence of a background magnetic field, thus  preserving \textit{a priori} the locality of the theory, we restrict our study to a non-commutative momentum space only; that is,
\[
\Theta_j=0,\qquad j=1, 2, 3. 
\]
Therefore, Eq. (\ref{transformation2}) is reduced to
\begin{eqnarray*}
\hat{r}_{i} & = & r_{i},\\
\hat{p}_{i} & = & p_{i}+\frac{\Omega_{ij}}{2}r_{j}=p_{i}-\frac{\left(\mathbf{\Omega}\times{\bf r}\right)_{i}}{2},
\end{eqnarray*}
where 
\begin{eqnarray*}
\left[\hat{r}_{i},\hat{r}_{j}\right] & = & 0,\\
\left[\hat{p}_{i},\hat{p}_{j}\right] & = & {\rm i}\Omega_{ij},\\
\left[\hat{r}_{i},\hat{p}_{j}\right] & = & {\rm i}\hbar\delta_{ij}.
\end{eqnarray*}

The comparison of the momenta commutation relations above with those obtained by the minimal coupling in the case of the presence a constant magnetic field \cite{Landau5} shows that the non-commutative parameters $\Omega_{ij}$ resemble the components of this magnetic field. From the previous equations and the properties of the Dirac matrices, we find that the Dirac equation (\ref{DiracCurved}) takes the explicit form:
\begin{eqnarray*}
m\Psi & = & \frac{i}{\sqrt{1-\omega^{2}\eta^{2}\rho^{2}}}\gamma^{0}\partial_{ t}\Psi+i\gamma^{2}\frac{\omega\eta\rho}{\sqrt{1-\omega^{2}\eta^{2}\rho^{2}}}\partial_{ t}\Psi\\
 &  & +i\gamma^{1}\left(\partial_{ \rho}-\frac{1}{2\rho}  +m\omega_{0}\gamma^{0}\rho-\frac{\Omega_{2}}{2i}z\right)\Psi\\
 &  & +i\frac{\sqrt{1-\omega^{2}\eta^{2}\rho^{2}}}{\eta\rho}\gamma^{2}\left(\partial_{ \varphi}-\frac{\Omega_{3}}{2i}\rho+\frac{\Omega_{1}}{2i}z\right)\Psi\\
 &  & +i\gamma^{3}\left(\partial_{ z}+\frac{\Omega_{2}}{2i}\rho+m\omega_{0}\gamma^{0}z\right)\Psi+\frac{\gamma^{0}}{2}\frac{\eta}{\left(1-\omega^{2}\eta^{2}\rho^{2}\right)}\omega\sigma^{3}\Psi,
\end{eqnarray*}
with the Pauli matrices given in Eq. (\ref{Pauli}).

If we write
\[
\Psi(t, \rho, \varphi, z)=e^{-i{\cal E}t}\left(\begin{array}{c}
\phi(\rho, \varphi, z)\\
\chi(\rho, \varphi, z)
\end{array}\right),
\]
we obtain
\beq
\begin{array}{l}
\left[{\cal E}-m\sqrt{1-\omega^{2}\eta^{2}\rho^{2}}+\frac{1}{2}\frac{\eta\omega}{\sqrt{1-\omega^{2}\eta^{2}\rho^{2}}}\sigma^{3}   \right]\phi  =  -i\sigma^{1}\sqrt{1-\omega^{2}\eta^{2}\rho^{2}}\left[\partial_{ \rho}-\frac{1}{2\rho}   -m\omega_{0}\rho+i\frac{\Omega_{2}}{2}z\right]\chi\\
    -i\sigma^{2}\frac{\left(1-\omega^{2}\eta^{2}\rho^{2}\right)}{\eta\rho}\left[\partial_{ \varphi}+i\frac{\Omega_{3}}{2}\rho-i\frac{\Omega_{1}}{2}z\right]\chi-\omega\eta\rho{\cal E}\sigma^{2}\chi
    -i\sigma^{3}\sqrt{1-\omega^{2}\eta^{2}\rho^{2}}\left[\partial_{ z}-zm\omega_{0}+i\frac{\Omega_{2}}{2}\rho\right]\chi
\end{array}\label{NC0A}
\eeq
and 
\beq
\begin{array}{l}
\left[{\cal E}+m\sqrt{1-\omega^{2}\eta^{2}\rho^{2}}+\frac{1}{2}\frac{\eta\omega}{\sqrt{1-\omega^{2}\eta^{2}\rho^{2}}}\sigma^{3}   \right]\chi  =  -i\sigma^{1}\sqrt{1-\omega^{2}\eta^{2}\rho^{2}}\left[\partial_{ \rho}-\frac{1}{2\rho}   +m\omega_{0}\rho+i\frac{\Omega_{2}}{2}z\right]\phi\\
  -i\sigma^{2}\frac{\left(1-\omega^{2}\eta^{2}\rho^{2}\right)}{\eta\rho}\left[\partial_{ \varphi}+i\frac{\Omega_{3}}{2}\rho-i\frac{\Omega_{1}}{2}z\right]\phi-\omega\eta\rho{\cal E}\sigma^{2}\phi
  -i\sigma^{3}\sqrt{1-\omega^{2}\eta^{2}\rho^{2}}\left[\partial_{ z}+zm\omega_{0}-i\frac{\Omega_{2}}{2}\rho\right]\phi.
\end{array}\label{NC0B}
\eeq
We point out two sign differences with Ref. \cite{Bakke} in each of these two equations: the third term on the left-hand side and, in the first line of the right-hand side, the second term  scaling with $\frac1{2\rho}$.

Hereafter, we write down the general equations with both $\rho$ and $z$ dependence in the oscillator. However later on, we will solve the simpler case with the oscillator depending only on $\rho$. We assume that the angular velocity of the rotating frame is small, so that $\omega\rho\ll 1$. Thus we can use the approximation $\sqrt{1-\omega^{2}\eta^{2}\rho^{2}}\approx 1-\frac12\omega^{2}\eta^{2}\rho^{2}$, and we can rewrite Eqs. (\ref{NC0A}) and (\ref{NC0B}) as
\beq
\begin{array}{l}
\left[{\cal E}-m\left(1-\frac{\omega^{2}\eta^{2}\rho^{2}}{2}\right)+ \frac{\eta\omega}{2}\left(1+\frac{\omega^{2}\eta^{2}\rho^{2}}{2}\right)\sigma^{3}\right]\phi  =  -i\sigma^{1}\left(1-\frac{\omega^{2}\eta^{2}\rho^{2}}{2}\right)\left[\partial_{\rho}- \frac{1}{2\rho}-m\omega_{0}\rho+i\frac{\Omega_{2}}{2}z\right]\chi\nonumber\\
  -i\sigma^{2}\left(\frac{1-\omega^{2}\eta^{2}\rho^{2}}{\eta\rho}\right)\left[\partial_{\varphi}+i\frac{\Omega_{3}}{2}\rho-i\frac{\Omega_{1}}{2}z\right]\chi-\omega\eta\rho{\cal E}\sigma^{2}\chi
 -i\sigma^{3}\left(1-\frac{\omega^{2}\eta^{2}\rho^{2}}{2}\right)\left[\partial_{z}-zm\omega_{0}+i\frac{\Omega_{2}}{2}\rho\right]\chi
\end{array}\label{NC1A}
\eeq
and
\beq
\begin{array}{l}
\left[{\cal E}+m\left(1-\frac{\omega^{2}\eta^{2}\rho^{2}}{2}\right)+ \frac{\eta\omega}{2}\left(1+\frac{\omega^{2}\eta^{2}\rho^{2}}{2}\right)\sigma^{3}\right]\chi  =  -i\sigma^{1}\left(1-\frac{\omega^{2}\eta^{2}\rho^{2}}{2}\right)\left[\partial_{\rho}- \frac{1}{2\rho}+m\omega_{0}\rho+i\frac{\Omega_{2}}{2}z\right]\phi\\
  -i\sigma^{2}\left(\frac{1-\omega^{2}\eta^{2}\rho^{2}}{\eta\rho}\right)\left[\partial_{\varphi}+i\frac{\Omega_{3}}{2}\rho-i\frac{\Omega_{1}}{2}z\right]\phi-\omega\eta\rho{\cal E}\sigma^{2}\phi
  -i\sigma^{3}\left(1-\frac{\omega^{2}\eta^{2}\rho^{2}}{2}\right)\left[\partial_{z}+zm\omega_{0}-i\frac{\Omega_{2}}{2}\rho\right]\phi.
\end{array}\label{NC1B}
\eeq
As done in Ref. \cite{Bakke}, we neglect 
\beq
\beta^2=\omega^{2}\eta^{2}\rho^{2}  \label{beta2}
\eeq 
in $1-\frac{\omega^{2}\eta^{2}\rho^{2}}{2}$, but we keep $\left(\frac{1-\omega^{2}\eta^{2}\rho^{2}}{\eta\rho}\right)$. The previous equations then become
\begin{eqnarray}\label{NC-1}
\left[{\cal E}-m+ \frac{\eta\omega}{2}\sigma^{3}\right]\phi & = & -i\sigma^{1}\left[\partial_{\rho}- \frac{1}{2\rho}-m\omega_{0}\rho+i\frac{\Omega_{2}}{2}z\right]\chi\nonumber\\
 &  & -i\sigma^{2}\left(\frac{1-\omega^{2}\eta^{2}\rho^{2}}{\eta\rho}\right)\left[\partial_{\varphi}+i\frac{\Omega_{3}}{2}\rho-i\frac{\Omega_{1}}{2}z\right]\chi-\omega\eta\rho{\cal E}\sigma^{2}\chi\nonumber\\
 &  & -i\sigma^{3}\left[\partial_{z}-m\omega_{0}z+i\frac{\Omega_{2}}{2}\rho\right]\chi
\end{eqnarray}
and
\begin{eqnarray}\label{NC-2}
\left[{\cal E}+m+ \frac{\eta\omega}{2}\sigma^{3}\right]\chi & = & -i\sigma^{1}\left[\partial_{\rho}- \frac{1}{2\rho}+m\omega_{0}\rho+i\frac{\Omega_{2}}{2}z\right]\phi\nonumber\\
 &  & -i\sigma^{2}\left(\frac{1-\omega^{2}\eta^{2}\rho^{2}}{\eta\rho}\right)\left[\partial_{\varphi}+i\frac{\Omega_{3}}{2}\rho-i\frac{\Omega_{1}}{2}z\right]\phi-\omega\eta\rho{\cal E}\sigma^{2}\phi\nonumber\\
 &  & -i\sigma^{3}\left[\partial_{z}+m\omega_{0}z-i\frac{\Omega_{2}}{2}\rho\right]\phi.
\end{eqnarray}

In order to obtain a differential equation for $\phi$, it is convenient to rewrite Eqs. (\ref{NC-1}) and (\ref{NC-2}) in a more concise form as 
\begin{equation}\label{NC-3}
T_+\phi=-i\sigma^1D_1^-\chi-i\sigma^2D_2\chi-i\sigma^3D_3^-\chi,
\end{equation} 
\begin{equation}\label{NC-4}
S_+\chi=-i\sigma^1D_1^+\phi-i\sigma^2D_2\phi-i\sigma^3D_3^+\phi,
\end{equation} 
where we defined
\begin{equation}\label{NC-S}
S_\pm\equiv {\cal E}+m\pm\frac 12\omega\eta\sigma^3,
\end{equation}
\begin{equation}\label{NC-T}
T_\pm\equiv {\cal E}-m\pm\frac 12\omega\eta\sigma^3,
\end{equation}
\begin{equation}\label{NC-D1}
D_1^\pm\equiv  \partial_{\rho}-\frac{1}{2\rho}\pm m\omega_{0}\rho+i\frac{\Omega_{2}}{2}z,
\end{equation}
\begin{equation}\label{NC-D2}
D_2\equiv \left(\frac{1-\omega^2\eta^2\rho^2}{\eta\rho}\right) \left[\partial_{\varphi}+i\frac{\Omega_{3}}{2}\rho-i\frac{\Omega_{1}}{2}z\right] -i\omega\eta\rho {\cal E}
\end{equation}
and
\begin{equation}\label{NC-D3}
D_3^\pm\equiv\partial_{z}\pm m\omega_{0}z\mp i\frac{\Omega_{2}}{2}\rho.
\end{equation}
Note the relations
\begin{equation}\label{S+S-}
S_+S_-=S_-S_+=\left({\cal E}+m\right)^2-\left(\frac 12\omega\eta\right)^2=:S^2,
\end{equation}
\begin{equation}\label{S-Pauli}
S_-\sigma^{1,2}=\sigma^{1,2}S_+,\qquad S_-\sigma^{3}=\sigma^{3}S_-,
\end{equation}
\begin{equation}\label{T-Pauli}
T_+\sigma^{1,2}=\sigma^{1,2}T_-,\qquad T_+\sigma^{3}=\sigma^{3}T_+,
\end{equation}
\begin{equation}\label{S+-2}
S_\pm^2=\left({\cal E}+m\right)^2+\left(\frac 12\omega\eta\right)^2\pm \left(E+m\right)\left(\omega\eta\sigma^3\right)
\end{equation}
and
\begin{equation}\label{T+S+}
S_+T_+=T_+S_+=\left({\cal E}+\frac 12\omega\eta\sigma^3\right)^2-m^2,
\end{equation}
which is analogous to the LHS of Eq (14) in Ref. \cite{Bakke}.

We can extract $\chi$ from Eq. (\ref{NC-4}) by multiplying it on the left with $S_-$. Then we find
\begin{equation}\label{S-xNC-4}
S_-S_+\chi=-i\sigma^1S_+D_1^+\phi-i\sigma^2S_+D_2\phi-i\sigma^3S_-D_3^+\phi,
\end{equation} 
which leads to
\begin{equation}\label{NC-chi}
\chi=-\frac i{S^2}\left(\sigma^1S_+D_1^+\phi+\sigma^2S_+D_2\phi+\sigma^3S_-D_3^+\phi\right).
\end{equation}

Next, let us multiply Eq. (\ref{NC-3}) on the left with $S_+$, so that
\begin{equation}\label{NC-S+T+}
S_+T_+\phi=-i(\sigma^1S_-D_1^-\chi+\sigma^2S_-D_2\chi+\sigma^3S_+D_3^-\chi).
\end{equation} 
These lead to
\begin{eqnarray}\label{Summary}
S_+T_+\phi&=&-D_1^-D_1^+\phi-D_2D_2\phi-D_3^-D_3^+\phi\nonumber\\
& & +i\sigma^1\left(D_3^-D_2\phi-D_2D_3^+\phi\right)\nonumber\\
& & +i\sigma^2\left(D_1^-D_3^+\phi-D_3^-D_1^+\phi\right)\nonumber\\
& & +i\sigma^3\left(D_2D_1^+\phi-D_1^-D_2\phi\right).
\end{eqnarray}

By calculating all the terms in Eq. (\ref{Summary}), we obtain
\begin{eqnarray}\label{Equation1}
\left({\cal E}+\frac    12\omega\eta\sigma^3\right)^2\phi-m^2\phi&=&m^2\omega_0^2\rho^2\phi +2i\omega{\cal E}\partial_\varphi\phi -\frac{3}{4\rho^2}\phi-m\omega_0\phi+\frac 1\rho\partial_\rho\phi \nonumber\\
& & -\frac 1{\eta^2\rho^2}\partial_{\varphi\varphi}\phi-\partial_{\rho\rho}\phi-\partial_{zz}\phi+2\omega^2\partial_{\varphi\varphi}\phi \nonumber\\
& & -i\Omega_2z\partial_\rho\phi+\frac{i}{\eta^2\rho^2}\Omega_1z\partial_\varphi\phi-\frac{i}{\eta^2\rho}\Omega_3\partial_\varphi\phi+\frac{i}{2\rho}\Omega_2z\phi  \nonumber\\
& & +2i\rho\omega^2\Omega_3\partial_\varphi\phi -2iz\omega^2\Omega_1\partial_\varphi\phi-\omega\rho{\cal E}\Omega_3\phi+\omega z{\cal E}\Omega_1\phi\nonumber\\
& & -\frac 12\eta\rho\omega^2\Omega_1 \sigma^1\phi+\frac{\Omega_1}{2\eta\rho} \sigma^1\phi+i\Omega_2\rho^2\omega\eta{\cal E}\sigma^1\phi-\frac{\Omega_2}\eta\sigma^1\partial_\varphi\phi\nonumber\\
& & -2im\omega_0\rho \sigma^2\partial_z\phi +\frac 12\Omega_2\sigma^2\phi+\Omega_2\rho\sigma^2\partial_\rho\phi\nonumber\\
& & +2\omega\eta m\omega_0\rho^2{\cal E}\sigma^3\phi+\frac i\eta 2  m\omega_0\sigma^3\partial_\varphi\phi-{\cal E}\omega\eta\sigma^3\phi +\frac i{\eta\rho^2}\sigma^3\partial_\varphi\phi+i\eta\omega^2\sigma^3\partial_\varphi\phi  \nonumber \\
& &   +\frac1{2\eta\rho^2}z\Omega_1\sigma^3\phi-\eta\omega^2\rho\Omega_3\sigma^3\phi +\frac12\eta\omega^2z\Omega_1\sigma^3\phi  \nonumber \\
& & -\frac\rho\eta m\omega_0\Omega_3\sigma^3\phi+\frac z\eta m\omega_0\Omega_1\sigma^3\phi+{\cal O}(\beta^2)+{\cal O}(\Omega^2),
\end{eqnarray}
where $\beta^2$ is defined in Eq. (\ref{beta2}). Among the commutative terms of Eq. (\ref{Equation1}), we observe five differences with Ref. \cite{Bakke}: the first sign of the LHS, and on the RHS, the third and fifth terms of line 1, the first and fourth terms of line 7. Therefore we should not expect our results to coincide, in their commutative limit, with the results of Ref. \cite{Bakke}.

To recapitulate how we obtained Eq. (\ref{Equation1}): we solved the generalized Dirac equation (\ref{DiracCurved}) with the tetrads of Eq. (\ref{Tetrads}) associated with the cosmic string metric in Eq. (\ref{CosmicString}), in a space that is non-commutative in momenta only, as described in Eq. (\ref{transformation2}) with ${\bf\Omega}={\bf 0}$. Finally, the oscillator is introduced via Eq. (\ref{Oscillator}) in the cylindrical coordinates of Eq. (\ref{Cylindrical}). In the following section, we shall consider further approximations in order to simplify and solve Eq. (\ref{Equation1}), and determine the energy eigenvalues. 




\section{Solutions of the radial oscillator with non-commutativity parallel to the string {\label{Solutions}}}

In this section, we solve the radial part of Eq. (\ref{Equation1}) by means of the ansatz in Eq. (\ref{phis}) with further simplifications: we keep only the non-commutativity around the axis of (frame) rotation non-zero, we consider solution in a plane perpendicular to the axis of rotation.  We utilize the Frobenius method to obtain solutions in terms of the biconfluent Heun functions and the energy eigenvalues in terms of the relevant parameters. 

If we write solutions in the form
\begin{equation}\label{phis}
\phi(\rho,\varphi,z)=e^{i\left( L+\frac12\right)\varphi}\left(
\begin{array}{c}
R_+(\rho,z)\\
R_-(\rho,z)\end{array}
\right),
\end{equation}
where $L=0,\pm 1, \pm 2, \dots$ \cite{Bakke}, which we substitute into Eq. (\ref{Equation1}) and collect terms  in $\rho^2$, $\frac 1{\rho^2}$, $1$, $\frac 1\rho\partial_\rho$, $\partial_{\rho\rho}$, $\frac1\rho$ and $\rho$, with each terms containing further dependences on $z$, we find
\beq
\begin{array}{l}\label{Equation3}
\left({\cal E}+\frac 12\omega\eta\sigma^3\right)^2\phi-m^2\phi= -\partial_{\rho\rho}\phi+\frac 1\rho\partial_\rho\phi -\partial_{zz}\phi -i\Omega_2z\partial_\rho\phi\nonumber\\
+\rho^2  \left[m^2\omega_0^2\phi +i\Omega_2\omega\eta{\cal E}\sigma^1\phi+2\omega\eta m\omega_0{\cal E}\sigma^3\phi\right]\nonumber\\ +\frac 1{\rho^2}  \left[-\frac1{\eta^2}\Omega_1z\left( L+\frac12\right)\phi -\frac 3{4}\phi+\frac 1{\eta^2}\left( L+\frac12\right)^2\phi -\frac 1{\eta}\sigma^3\left( L+\frac12\right)\phi+\frac1{2\eta}z\Omega_1\sigma^3\phi\right]\nonumber\\
  + \left[-2\omega{\cal E}\left( L+\frac12\right)\phi-m\omega_0\phi-2\omega^2\left( L+\frac12\right)^2\phi+2z\omega^2\Omega_1\left( L+\frac12\right)\phi\right.\nonumber\\
  \qquad\left. +\omega z{\cal E}\Omega_1\phi -\frac{i\Omega_2}\eta\sigma^1\left( L+\frac12\right)\phi+\frac 12\Omega_2\sigma^2\phi -\frac 1\eta 2  m\omega_0\sigma^3\left( L+\frac12\right)\phi-{\cal E}\omega\eta\sigma^3\phi \right.\nonumber\\
  \qquad \left. -\eta\omega^2\sigma^3\left( L+\frac12\right)\phi +\frac12\eta\omega^2z\Omega_1\sigma^3\phi +\frac z\eta m\omega_0\Omega_1\sigma^3\phi   \right]\nonumber\\
  +\frac1\rho   \left[\frac{1}{\eta^2}\Omega_3\left( L+\frac12\right)\phi+\frac{i}{2}\Omega_2z\phi+\frac{\Omega_1}{2\eta} \sigma^1\phi\right] \nonumber\\
  +\rho  \left[-2\omega^2\Omega_3\left( L+\frac12\right)\phi  -\omega{\cal E}\Omega_3\phi  -\frac 12\eta\omega^2\Omega_1 \sigma^1\phi -2im\omega_0\sigma^2\partial_z\phi+\Omega_2\sigma^2\partial_\rho\phi\right.\nonumber\\
  \qquad \left. -\eta\omega^2\Omega_3\sigma^3\phi -\frac1\eta m\omega_0\Omega_3\sigma^3\phi\right]+{\cal O}(\beta^2)+{\cal O}(\Omega^2).
\end{array}
\eeq

Now let us examine  Eq. (\ref{Equation3}) with $\Omega_1=0=\Omega_2$, so that only the non-commutativity parameter $\Omega_3$,  parallel to the cosmic string, is different from zero.  In this case, the right-hand side of Eq. (\ref{Equation3}) commutes with $p_z=-i\partial_z$ so that we can use
\begin{equation}\label{phis2}
\phi_s(\rho,\varphi,z)=e^{i\left( L+\frac12\right)\varphi}e^{im_\ell z}\left(
\begin{array}{c}
R_+(\rho)\\
R_-(\rho)\end{array}
\right).
\end{equation}
If we set $m_\ell=0$ so that we work with a planar system, then we can express Eq. (\ref{Equation3}) in terms of eigenfunctions of $\sigma^3$ with $s=\pm1$ corresponding to $R_+(\rho)$ and $R_-(\rho)$,  (where prime denotes differentiation with respect to $\rho$)
\beq
\begin{array}{l}\label{Equation5}
\left({\cal E}^2+s{\cal E}\omega\eta+\frac14\omega^2\eta^2 -m^2\right)R_s = -R_s''+\frac 1\rho R_s'+ \rho^2  \left[m^2\omega_0^2 +2\omega\eta m\omega_0{\cal E}s\right]R_s\nonumber\\
  +\frac 1{\rho^2}  \left[ -\frac 3{4}+\frac 1{\eta^2}\left( L+\frac12\right)^2 -\frac s{\eta}\left( L+\frac12\right)\right]R_s\nonumber\\
  + \left[-2\omega{\cal E}\left( L+\frac12\right)-m\omega_0-2\omega^2\left( L+\frac12\right)^2 \right.\nonumber\\
  \qquad\left.  -\frac 1\eta 2  m\omega_0s\left( L+\frac12\right)-{\cal E}\omega\eta s -\eta\omega^2s\left( L+\frac12\right)   \right]R_s 
+\frac{\Omega_3}{\rho\eta^2} \left( L+\frac12\right)R_s\nonumber\\
  +\rho  \left[-2\omega^2\Omega_3\left( L+\frac12\right)  -\omega{\cal E}\Omega_3 -\eta\omega^2\Omega_3 s -\frac1\eta m\omega_0\Omega_3s\right]R_s.
\end{array}\eeq

Let us write Eq. (\ref{Equation5}) in a more compact form, 
\begin{equation}\label{Equation6}
R_s''-\frac 1\rho R_s'+AR_s+B\rho^2R_s+\frac {C}{\rho^2}R_s+D\rho R_s+\frac E\rho R_s=0,
\end{equation} 
where 
\begin{equation}\label{defA}
\begin{array}{lcl}
A&\equiv&{\cal E}^2+s{\cal E}\omega\eta+\frac14\omega^2\eta^2 -m^2 +\left[\frac{m\omega_0}\eta+s\omega{\cal E}+s\omega^2\left(L+\frac 12\right)\right]
\left[\eta+2s\left(L+\frac 12\right)\right],
\end{array}
\end{equation}
\begin{equation}\label{defB}
B\equiv -m^2\omega_0^2 -2\omega\eta m\omega_0{\cal E}s,
\end{equation}
\begin{equation}\label{defC}
C\equiv \frac 3{4}-\frac 1{\eta^2}\left( L+\frac12\right)^2 +\frac s{\eta}\left( L+\frac12\right),
\end{equation}
\begin{equation}\label{defD}
D\equiv 2\omega^2\Omega_3\left( L+\frac12\right)  +\omega{\cal E}\Omega_3 +\eta\omega^2\Omega_3 s +\frac1\eta m\omega_0\Omega_3s,
\end{equation}
\begin{equation}\label{defE}
E\equiv -\frac{\Omega_3}{\eta^2} \left( L+\frac12\right).
\end{equation}


In order to investigate the energy spectrum, let us change the dependent variable as
\beq
R_s(\rho)=e^{-k\rho^{2}}e^{-q\rho}\rho^{h}F\left(\rho\right),
\label{ChangeVariables}\eeq
with $k$, $q$, $h$ to be yet determined, such that by substituting it in Eq. (\ref{Equation6}), we find
\begin{eqnarray}
& & F^{\prime\prime}\left(\rho\right) +\left(-4k\rho+\frac{2h-1}{\rho}-2q\right)F^{\prime}\left(\rho\right)  +\left[\frac{E+q\left(1-2h\right)}{\rho}+A+q^{2}-4kh\right.\nonumber\\
&  & +\left.\left(4k^{2}+B\right)\rho^{2}+\left(D+4kq\right)\rho+\frac{C+h\left(h-2\right)}{\rho^{2}}\right]F\left(\rho\right)=0.\label{Fpp}
\end{eqnarray}
 The terms proportional  to $\rho^2$, $\rho$ and $\frac 1{\rho^2}$ in Eq. (\ref{Fpp}) are set equal to zero if we choose $k$, $q$ and $h$ in Eq. (\ref{ChangeVariables}) as follows: 
\beq\label{Def-k}
4k^{2}+B=0\Rightarrow k=\pm\frac{\sqrt{-B}}{2},
\eeq
\beq\label{Def-q}
4kq+D=0\Rightarrow q=-\frac{D}{2\left(\pm\sqrt{-B}\right)},
\eeq
\beq\label{Def-h}
C+h\left(h-2\right)=0\Rightarrow h=1\pm\sqrt{1-C}.
\eeq
Then, Eq. (\ref{Fpp}) reduces to
\beq\label{Equation7}
\rho F^{\prime\prime}\left(\rho\right) +\left[2h-1-2q\rho-4k\rho^{2}\right]F^{\prime}\left(\rho\right)  +\left[E+q\left(1-2h\right)+\left(A+q^{2}-4hk\right)\rho\right]F\left(\rho\right) =0.
\eeq
This can be compared to the biconfluent Heun equation in its canonical form \cite{Decarreau,Ronveaux},
\beq
zu^{\prime\prime}(z)+\left(1+\alpha-\beta z-2z^2\right)u^\prime(z)+\left[(\gamma-\alpha-2)z-\frac12\left[\delta+(1+\alpha)\beta\right]\right]u(z)=0.
\label{Heun}
\eeq

An interesting paper that examines the Klein-Gordon and Dirac equations of particles in cosmic string space-time in the presence of magnetic field and scalar potential, and which explains the use of the biconfluent Heun differential equation in the present context is Ref. \cite{Medeiros}: its appendix describes some properties of the Heun equation with the Frobenius method, and its introduction mentions references where Heun functions appear in quantum field theory. Other papers which involve the biconfluent Heun function in the context of quantum fields in the presence of topological defects include Refs. \cite{Ref88, Ref89}. 

We determine the appropriate sign for $k$ in Eq. (\ref{Def-k}) as follows. If we claim that the non-rotating frame solution should be obtained as a particular case of our solution, then with $\omega\rightarrow 0$ we have $0<\rho<\infty$. In this case, in the limit $\rho\rightarrow\infty$, Eq. (\ref{Equation7}) becomes
\beq\label{LargeRho}
0  \approx  F^{\prime\prime}\left(\rho\right)-4k\rho F^{\prime}\left(\rho\right)+\left(A+q^{2}-4hk\right)F\left(\rho\right).
\eeq
This limit suggests that $F\left(\rho\right)$ will behave like
\[
F\left(\rho\right)\approx e^{-\gamma\rho^{2}}\rho^{\lambda},
\]
so that by substituting into Eq. (\ref{LargeRho}) and keeping only the higher contributions, we find
\[
0  \approx  4\gamma^{2}e^{-\gamma\rho^{2}}\rho^{\lambda+2}+8k\gamma e^{-\gamma\rho^{2}}\rho^{\lambda+2} \approx  \left(\gamma+2k\right)4\gamma e^{-\gamma\rho^{2}}\rho^{\lambda+2},
\]
which implies that $\gamma=-2k$ and
\(
F\left(\rho\right)\approx e^{2k\rho^{2}}\rho^{\lambda},
\)
thus showing that  $k>0$ would imply that if $F(\rho)$ is divergent then $R(\rho)$ would diverge as well. We will avoid divergence of $R(\rho)$ by choosing $k>0$ but we enforce $F(\rho)$ to be be a finite polynomial, so that it does not diverge.  Therefore, Eq. (\ref{Def-k}) should be restricted to
\beq
k=+\frac{\sqrt{\left|B\right|}}{2},
\label{Def2-k}\eeq
where $\left|B\right|=-B$ since $B$ is negative. Consequently, Eq. (\ref{Def-q}) should be reduced to
\beq\label{Def2-q}
q=\frac{-D}{2\sqrt{\left|B\right|}}.
\eeq


Once we have established the appropriate sign of $k$ using the non-rotating limit, we return to the general case with an arbitrary $\omega$. The rotating frame imposes a hard-wall condition, i.e. $R(\rho\geq\rho_{0})=0$, where $\rho_{0}\equiv \frac{1}{\omega\eta}$ \cite{Bakke}. According to the ansatz given in Eq. (\ref{ChangeVariables}), we should enforce $F(\rho_{0})=0$. This condition is not attainable exactly for the polynomials that will be obtained below; that is, we cannot find an analytical form that strictly satisfies the hard-wall condition. This fact is consistent with a statement in Ref. \cite{ValentimBakke} for harmonic oscillators with a hard-wall confinement. However, it suffices that the amplitude of the wave function be negligibly small at distances larger than the hard-wall. In Appendix \ref{ApB} we demand the hard-wall condition is indeed satisfied and obtain Eq. (\ref{hardwallconstr}) which provides a constraint upon the set of the physical parameters in terms of any desired statistical significance (in terms of a coverage factor $j$) \cite{StatBook}. In other words, Eq. (\ref{hardwallconstr}) allows to constrain the physical parameters so that the probability of finding a particle will be within the desired confidence level respecting the hard-wall condition.


We now turn to  power series solutions of Eq. (\ref{Equation7}) by means of the Frobenius method. We consider
\begin{equation}
F(\rho) = \sum_{n=0}c_{n}\rho^{n+r}.
\label{PowerSeries}
\end{equation}
By substituting $F(\rho)$ into Eq. (\ref{Equation7}), we obtain
\begin{eqnarray*}
 &  & r\left(r-2+2h\right)c_{0}\rho^{-1+r}\\
 &  & +\left\{ \left(1+r\right)\left(r+2h-1\right)c_{1}+\left[-2qr+E+q\left(1-2h\right)\right]c_{0}\right\} \rho^{r}\\
 &  & +\sum_{n=0}\left(n+2+r\right)\left(n+r+2h\right)c_{n+2}\rho^{n+1+r}\\
 &  & +\sum_{n=0}\left[-2q\left(n+1+r\right)+E+q\left(1-2h\right)\right]c_{n+1}\rho^{n+1+r}\\
 &  & +\sum_{n=0}\left[-4k\left(n+r\right)+A+q^{2}-4kh\right]c_{n}\rho^{n+1+r}=0.
\end{eqnarray*}
From the coefficients of $\rho^{-1+r}$, we find, using $h$ from Eq. (\ref{Def-h}),
\begin{equation}
r=0\ {\mathrm{or}}\ 
r=2(1-h) =\pm 2\sqrt{1-C}.\label{def-r}
\end{equation}
For $\rho^r$, we obtain
\begin{equation}
c_{1}=\frac{2qr-E-q\left(1-2h\right)}{\left(1+r\right)\left(r+2h-1\right)}c_{0},\label{after55}
\end{equation}
and the coefficients of $ \rho^{n+1+r}$ lead to
\begin{eqnarray}
c_{n+2} & = & \frac{2q\left(n+1+r\right)-E-q\left(1-2h\right)}{\left(n+2+r\right)\left(n+r+2h\right)}c_{n+1}\nonumber\\
 &  & +\frac{4k\left(n+r\right)-A-q^{2}+4kh}{\left(n+2+r\right)\left(n+r+2h\right)}c_{n}.\label{cn+2}
\end{eqnarray}

The power series becomes a polynomial if there is a value of $n$, say $n_{0}$, such that the numerator of the last term in Eq. (\ref{cn+2}) is zero,
\begin{eqnarray}
A&=&4k\left(n_{0}+r+h\right)-q^{2},\label{conditionA}\\
c_{n_0+1} & = & 0.\label{conditioncn+1}
\end{eqnarray}
The first condition leads to
\begin{eqnarray}
&{\cal E}^{2}+\frac{\omega^{2}\eta^{2}}{4}+s{\cal E}\omega\eta-m^{2}+\left[\frac{m\omega_{0}}{\eta}+s\omega{\cal E}+s\omega^{2}\left(L+\frac{1}{2}\right)\right]\left[\eta+2s\left(L+\frac{1}{2}\right)\right]\nonumber\\
&=4k\left(n_{0}+r+h\right)-q^{2}.\label{eq59}
\end{eqnarray}
Note that, although this equation provides the energy spectrum, the condition $c_{n_0+1}=0$ in Eq. (\ref{conditioncn+1}) imposes further constraints between physical parameters describing our system. In our situation, there is no general expression for the biconfluent Heun function for general $n_0$ and since we cannot obtain a solution of Eq. (\ref{conditioncn+1}) for a general $n_0$, we are left to solve it for each degree of the polynomial (for an early example, see Refs. \cite{Vercin1,Vercin2}) considered one at a time. For instance, we show in the Appendix A that the condition $c_2=0$, which corresponds to a linear polynomial, can be satisfied by adjusting the oscillator's frequency $\omega_0$ in the limit $\frac{\omega\eta}{m}\ll1$, as follows:
\begin{equation}
\omega_{0} \approx-\frac{\omega\eta}{m}s{\cal E}_{n_0=1}+\frac{K}{m},\label{omega0-c2=0}
\end{equation}
where
\[
K\equiv \frac{1}{2}\left(Z^{2}-\frac{s}{\eta}Z\right)\frac{\Omega_{3}^{2}}{\left(1+r\right)\left(r+2h-1\right)} 
\]
and
\begin{align*}
Z\equiv \frac{s}{2\eta}\left[1-2\left(r+h\right)\right]+\frac{1}{\eta^{2}}\left(L+\frac{1}{2}\right) .
\end{align*}

From Eqs. (\ref{def-r}), (\ref{Def-h}) and (\ref{defC}),
\begin{equation}
r+h=1+\left|\frac12-\frac s\eta\left(\frac12+L\right)\right|,
\end{equation}
We chose $r=2\left(1-h\right)$ because by analyzing the commutative limit, we observe that we must keep the minus sign in Eqs. (\ref{Def-h}) and the plus sign in Eq. (\ref{def-r}) if we want our results to be compatible. (With $r=0$, the minus sign should be discarded and the plus sign should be considered.) By substituting $\omega_{0}$ in Eq. (\ref{eq59}), with $q$ from Eqs. (\ref{Def2-q}),  (\ref{defB}) and (\ref{defD}), we obtain in first order of $\frac{\omega\eta}{m}$,:
\begin{align}
0\approx	\left(\frac{{\cal E}}{m}\right)^{2}+s\left(\frac{{\cal E}}{m}\right)\frac{\omega\eta}{m}-1+4\frac{K}{m^{2}}\left[-1+\frac{s}{\eta}\left(L+\frac{1}{2}\right)\right]+\frac{\Omega_{3}^{2}}{m^{2}}\frac{1}{4\eta^{2}}.
\end{align}
This equation has the following solution:
\begin{align}
\frac{{\cal E}_{n_0=1}}{m}\approx-\frac{s}{2}\frac{\omega\eta}{m}\pm\sqrt{1-\frac{\Omega_{3}^{2}}{m^{2}}\frac{1}{4\eta^{2}}-4\frac{K}{m^{2}}\left[-1+\frac{s}{\eta}\left(L+\frac{1}{2}\right)\right]} \label{finalE}
\end{align}
As mentioned earlier, Eq. (\ref{eq59}) provides the energy spectrum whereas Eq. (\ref{finalE}) was obtained by enforcing the condition (\ref{conditionA}) for the mode $n_0=1$; thus Eq. (\ref{finalE}) is valid for $n_0=1$ only.

Let us mention two physical observations:  (1) there is a coupling between the angular momentum, represented by the quantum number $L$, the spin (quantum number $s$) and the non-commutativity parameter $\Omega_{3}$ in the last term on the right-hand side and (2) there is a coupling between the spin and the angular velocity $\omega$ is shown in the first term on the right-hand side. The coupling between the angular momentum and the angular velocity $\omega$ is a second order effect. 

The energy obtained in Eq. (\ref{finalE}) should also be consistent with the hard-wall condition. We have shown in Appendix \ref{ApB} that this is essentially the case  if the following constraint is satisfied: $\frac{\Omega_{3}}{\omega}\gtrsim j$, where $j$ is the coverage factor. We obtain expressions in terms of the physical parameters that provide the desired statistical significance related to the coverage factor. The larger the value of $j$, the greater the probability density  within the hard-wall $\rho_0=\frac{1}{\eta\omega}$. Therefore, there is a compromise between the strength of the non-commutative parameter $\Omega_3$ and the angular velocity of the rotating frame $\omega$. In any case, this requirement is in agreement with $\frac{\omega\eta}{m}\ll 1$. Next, we consider in turn the limits where the angular velocity $\omega$ approaches zero, the large-mass limit, and the commutative limit $\Omega_3\approx 0$.

\paragraph*{Low angular velocity limit $\omega\rightarrow0$}

In this limit, Eq. (\ref{finalE}) reduces to
\begin{align}
\frac{{\cal E}_{n_0=1}}{m}\approx\pm\sqrt{1-\frac{\Omega_{3}^{2}}{m^{2}}\frac{1}{4\eta^{2}}-4\frac{K}{m^{2}}\left[-1+\frac{s}{\eta}\left(L+\frac{1}{2}\right)\right]}, \label{low-omega}
\end{align}
(with $K$ defined after Eq. (\ref{omega0-c2=0}))
where the contribution of non-commutativity is described by the second and third terms with the factor $\Omega_{3}^{2}$, inversely proportional to $m^2\eta^2$; thus we observe that the bigger the mass the smaller the contribution of the term with the non-commutative parameter. (We emphasize again that Eq. (\ref{finalE}) corresponds only to the mode $n_0=1$, thus so does Eq. (\ref{low-omega}).) It is interesting to note that 
\begin{equation}
Z =\begin{cases}
0, & s\left[s-\frac{2}{\eta}\left(L+\frac{1}{2}\right)\right]\le0,\\
-\frac{s}{\eta}\left|s-\frac{2}{\eta}\left(L+\frac{1}{2}\right)\right|, & s\left[s-\frac{2}{\eta}\left(L+\frac{1}{2}\right)\right]\ge0,
\end{cases}
\end{equation}
where $Z$ is defined in Eq. (\ref{defZ}) of the Appendix.
In the first case, $K=0$ and the contribution of the non-commutative parameter decreases the modulus of the energy. The contribution of the non-commutative parameter is similar in the second case, where $K\neq0$: the contribution of the non-commutative terms, that now accounts for both the second and third terms of the right hand side, also makes the modulus of the energy decrease.

\paragraph*{Large-mass limit}

With the same reasoning as in Eqs. (22) and (26) of Ref. \cite{Bakke}, we consider the case in which the mass $m$ is greater than the other physical parameters in the energy $\mathcal{E}_{n_0=1}$. In this limit, all the terms in the square root of Eq.~(\ref{finalE}) other than 1 are small compared to 1, which allows us to apply the approximation $\sqrt{1+x}\approx1+\frac{1}{2}x$. Then Eq.~(\ref{finalE}) becomes 
\begin{align}
\frac{{\cal E}_{n_0=1}}{m}\approx\pm\left(1-\frac{1}{8\eta^{2}}\frac{\Omega_{3}^{2}}{m^{2}}-2\frac{K}{m^{2}}\left[-1+\frac{s}{\eta}\left(L+\frac{1}{2}\right)\right]\right)-\frac{s}{2}\frac{\omega\eta}{m} \, .
 \label{nonrel}
\end{align}

The first term on the right hand side of Eq. (\ref{nonrel}) represents the rest energy of the Dirac particle, whereas the remaining terms give the contribution to the energy of a Dirac oscillator subjected to non-inertial effects of the rotating frame and the topology of the cosmic string in a space that is non-commutative along the string. As in the previous case, the contribution of the non-commutative terms makes the modulus of the energy decrease. The contribution of the frame angular velocity term imply the increasing or decreasing of the energy, depending on the value of the spin.

\paragraph*{Commutative case}

For the relativistic case, we consider the limit $\Omega_{3}\rightarrow 0$ in Eq. (\ref{finalE}). This leads to

\begin{align*}
\frac{{\cal E}_{n_0=1}}{m}\approx\pm1-\frac{s}{2}\frac{\omega\eta}{m}.
\end{align*}
It may suggest that the energy is simply the rest energy with a contribution due to the frame rotation but one must be careful when considering this commutative limit.  As can be seen from Ref. \cite{Bakke}, when one begins from a commutative space, then the condition $c_{n_{0}+1}=0$ need not be imposed, as the wave function is not related to the Heun functions. Without this condition, $\omega_{0}$ would not acquire discrete values and would remain as a free parameter of the theory. 

Eq. (\ref{finalE}) should be considered with $q=0$, which would lead to
\begin{equation}
\frac{{\cal E}_{n_0=1}}{m}\approx\begin{cases}
\frac{1}{2}s\frac{\omega\eta}{m}\pm\sqrt{1+2\frac{\omega_{0}}{m}}, & \left|\frac{1}{2}-\frac{s}{\eta}\left(\frac{1}{2}+L\right)\right|\le0\\
s\frac{\omega\eta}{m}\left[\frac{3}{2}-\frac{2s}{\eta}\left(\frac{1}{2}+L\right)\right]\pm\sqrt{1-4\frac{\omega_{0}}{m}\left[-1+\frac{s}{\eta}\left(L+\frac{1}{2}\right)\right]}, & \left|\frac{1}{2}-\frac{s}{\eta}\left(\frac{1}{2}+L\right)\right|>0
\end{cases}
\end{equation}
In this case, we observe that the coupling between $\omega$ and the angular momentum is present in the first order approximation when $\left|\frac{1}{2}-\frac{s}{\eta}\left(\frac{1}{2}+L\right)\right|>0$.




\section{Final remarks}

This paper analyzes the non-commutative effects on a Dirac oscillator in a cosmic string space-time equipped with a rotating frame.  The non-commutativity is introduced only in the momentum coordinates, since the intention was to emulate a background magnetic field. The non-commutativity is a new feature compared to similar studies in the literature, as in Ref.~\cite{Bakke}.

We build the field equation for the spinor particle in the curved space-time using the tetrad formalism. The Dirac oscillator is introduced via a non-minimal coupling to the momentum; the oscillations with frequency $\omega_0$ are, at first, allowed to take place in both $\rho$- and $z$-direction of a cylindrical coordinate system appropriate to the geometry of the string. Later, when solving for the self-energy of the Dirac particle, the oscillations are restricted to the radial direction (perpendicular to the string).

In order to decouple the field equations for the two components of the spinor, we make the assumption that the rotating frame velocity $\omega \rho$ is much smaller than the speed of light $c=1$. The mass $m$ of the Dirac particle is also taken to be greater than the parameter $\omega \eta $ related to the effects of the rotating frame's angular speed $\omega$ and parameter $\eta$ associated to the string's mass density.

The decoupled field equation is solved in terms of eigenfunctions of the Dirac matrix $\sigma^3$ under the restriction that the momentum non-commutativity lies along the string axis. Accordingly, the presence of non-commutativity is controlled solely by parameter $\Omega_3$. The differential equation for the radial part of the wave function is obtained through Frobenius method and corresponds to the biconfluent Heun equation. Physically meaningful eigenfunctions require the energy $\mathcal{E}$ to be discretized. Besides the discretization equation, Eq. (\ref{conditionA}), we must also satisfy condition $c_{n_0+1}=0$ for the series termination. This last requirement prevented us from obtaining a closed expression for the energy spectrum $\mathcal{E}$ in terms an arbitrary value of $n_0$. This has to be done order-by-order for each value chosen for $n_0$. As an example, we set $n_0=1$ and obtained the analytic formula for  $\mathcal{E}_{n_0=1}$. On top of that, the hard-wall condition was imposed. We found the restrictions on the physical parameter necessary to make the probability density of finding the particle outside the hard-wall negligibly small. Appendix B makes it clear that the non-commutativity parameter plays a role for implementing the hard-wall confining condition.

Hereafter we stress the novelties coming from non-commutativity. The non-commutative parameter $\Omega_3$ couples to orbital angular momentum $L$, to spin $s$ and to the parameters related to frame rotation frequency $\omega$, string's mass density related quantity $\eta$ and mass $m$ of the particle. This is clear from e.g. Eq. (\ref{finalE}). This equation also shows how $\Omega_3$ modifies the energy of the Dirac oscillator:  the modulus of the energy decreases due to the non-commutative parameter. The non-commutative term scales with the inverse of the mass: the bigger the mass, the smaller the contribution of the NC term.

The non-commutative term in the various expressions of $\mathcal{E}$ scales with inverse powers of $\eta = 1-4\Lambda$, where $\Lambda$ is the string's mass density. Therefore, the bigger the density, the smaller the $\eta$ and the bigger the influence of the non-commutativity on the energy. However, the density modifies the contribution of the non-commutative term up to the limit $\eta=1$, when the conic space-time geometry produced by the string flattens out to Minkowski space-time leaving behind only the information about the Dirac oscillator.

Future perspectives of this line of research include the analysis of the DKP field on a non-commutative cosmic string space-time with a rotating frame.




\section*{Acknowledgements}

MdeM is grateful to the Natural Sciences and Engineering Research Council (NSERC) of Canada for partial financial support (grant number RGPIN-2016-04309), and to the Instituto Tecnol\'ogico de Aeron\'autica (SP, Brazil) and the Physics Department, McGill University (Montreal, Canada) for their hospitality.  RRC thanks CAPES-Brazil (grant number 88881.119228/2016-01) for partial financial support.  He is also grateful to the University of Alberta (Edmonton, Canada) and Instituto Tecnol\'ogico de Aeron\'autica (SP, Brazil) for their hospitality. The authors thank the anonymous referees who helped them to improve the paper.

\appendix


\renewcommand{\theequation}{A.\arabic{equation}}

\setcounter{equation}{0}

\section{Appendix: Restriction from the power series to a polynomial \label{ApA} }

The purpose of this Appendix is to show some details of the analysis of Eq.(\ref{conditioncn+1}). It leads immediately to 
\beq\label{fromConds}
\left\{ q\left[1-2\left(n_{0}+r+h\right)\right]+E\right\} c_{n_{0}}+4kc_{n_{0}-1} = 0.
\eeq
Hereafter, we consider the approximation 
\beq\label{eta-omega-m}
\frac{\omega\eta}{m}\ll1,
\eeq

The approximation (\ref{eta-omega-m}) also implies that Eq. (\ref{Def2-q}) gives
\[
q \approx-\frac{\omega\eta}{m}\frac{\omega}{2\omega_{0}}\left[s+\frac{2}{\eta}\left(L+\frac{1}{2}\right)\right]\Omega_{3}-\frac{s\Omega_{3}}{2\eta},
\]
from Eq. (\ref{Def2-k}) we find
\[
k\approx\frac{m\omega_{0}}{2}\left(1+\frac{s{\cal E}}{\omega_{0}}\frac{\eta\omega}{m}\right),
\]
and from Eqs. (\ref{Def-h}) and (\ref{def-r}), 
\[
h+r=1+\frac{1}{2}\left|s-\frac{2}{\eta}\left(L+\frac{1}{2}\right)\right|.
\]
With these parameters, we find from Eq. (\ref{fromConds}) that
\begin{eqnarray*}
0 & = & -\left\{ \left(-\frac{\omega\eta}{m}\frac{\omega}{2\omega_{0}}\left[s+\frac{2}{\eta}\left(L+\frac{1}{2}\right)\right]\Omega_{3}-\frac{s\Omega_{3}}{2\eta}\right)\left[1-2\left(n_{0}+r+h\right)\right]-\frac{\Omega_{3}}{\eta^{2}}\left(L+\frac{1}{2}\right)\right\} c_{n_{0}}\\
 &  & -4\frac{m\omega_{0}}{2}\left(1+\frac{s{\cal E}}{\omega_{0}}\frac{\eta\omega}{m}\right)c_{n_{0}-1}.
\end{eqnarray*}

We consider $n_0=1$ so that the Heun power series is simply restricted to a linear polynomial. Then
\begin{align*}
c_{1} & =\frac{2qr-E-q\left(1-2h\right)}{\left(1+r\right)\left(r+2h-1\right)}c_{0}\\
 & \approx\frac{\frac{1}{\eta^{2}}\left(L+\frac{1}{2}\right)+\left[\frac{\omega\eta}{m}\frac{\omega}{2\omega_{0}}\left[s+\frac{2}{\eta}\left(L+\frac{1}{2}\right)\right]+\frac{s}{2\eta}\right]\left[1-2\left(r+h\right)\right]}{\left(1+r\right)\left(r+2h-1\right)}\Omega_{3}c_{0},
\end{align*}
and
\begin{eqnarray*}
0 & = & -\left\{ \left(-\frac{\omega\eta}{m}\frac{\omega}{2\omega_{0}}\left[s+\frac{2}{\eta}\left(L+\frac{1}{2}\right)\right]-\frac{s}{2\eta}\right)\left[1-2\left(1+r+h\right)\right]-\frac{1}{\eta^{2}}\left(L+\frac{1}{2}\right)\right\} \times\\
 &  & \times\frac{\frac{1}{\eta^{2}}\left(L+\frac{1}{2}\right)+\left[\frac{\omega\eta}{m}\frac{\omega}{2\omega_{0}}\left[s+\frac{2}{\eta}\left(L+\frac{1}{2}\right)\right]+\frac{s}{2\eta}\right]\left[1-2\left(r+h\right)\right]}{\left(1+r\right)\left(r+2h-1\right)}\Omega_{3}^{2}c_{0}\\
 &  & -4\frac{m\omega_{0}}{2}\left(1+\frac{s{\cal E}}{\omega_{0}}\frac{\eta\omega}{m}\right)c_{0}.
\end{eqnarray*}
We define
\[
X\equiv\frac{\omega}{2\omega_{0}}\left[s+\frac{2}{\eta}\left(L+\frac{1}{2}\right)\right],
\] 
and we can write the previous equation as
\begin{eqnarray*}
0 & = & \frac{\left[\left(\frac{\omega\eta}{m}X+\frac{s}{2\eta}\right)\left[1-2\left(r+h\right)\right]+\frac{1}{\eta^{2}}\left(L+\frac{1}{2}\right)\right]^{2}}{\left(1+r\right)\left(r+2h-1\right)}\Omega_{3}^{2}\\
 &  & -2\left(\frac{\omega\eta}{m}X+\frac{s}{2\eta}\right)\frac{\left[\left(\frac{\omega\eta}{m}X+\frac{s}{2\eta}\right)\left[1-2\left(r+h\right)\right]+\frac{1}{\eta^{2}}\left(L+\frac{1}{2}\right)\right]}{\left(1+r\right)\left(r+2h-1\right)}\Omega_{3}^{2}
 -4\frac{m\omega_{0}}{2}\left(1+\frac{s{\cal E}}{\omega_{0}}\frac{\eta\omega}{m}\right).\end{eqnarray*}
We define $Y$ as the term that is proportional to $\Omega_{3}^2$, so that
\begin{eqnarray*}
0 & = & Y\Omega_{3}^{2}-4\frac{m\omega_{0}}{2}\left(1+\frac{s{\cal E}}{\omega_{0}}\frac{\eta\omega}{m}\right).
\end{eqnarray*}
In first order of $\frac{\omega\eta}{m}$, $Y$ can be rewritten as 
\begin{align*}
\left(1+r\right)\left(r+2h-1\right)Y & =\frac{\omega\eta}{m}\frac{s}{\eta}XW+Z^{2}-\frac{s}{\eta}Z,
\end{align*}
where we defined
\beq\label{defZ}
Z\equiv\frac{s}{2\eta}\left[1-2\left(r+h\right)\right]+\frac{1}{\eta^{2}}\left(L+\frac{1}{2}\right)
\eeq
and
\[
W\equiv4\left(r+h\right)^{2}-4\left(r+h\right)\frac{s}{\eta}\left(L+\frac{1}{2}\right)-1.
\]
We can express the previous equation as
\beq\label{TBS2}
\begin{aligned}
& \frac{\omega\eta}{m}s\left[\frac{1}{\eta}\frac{\omega}{2\omega_{0}}\left[s+\frac{2}{\eta}\left(L+\frac{1}{2}\right)\right]W\frac{\Omega_{3}^{2}}{\left(1+r\right)\left(r+2h-1\right)}-2m{\cal E}\right]\\
  & +\left(Z^{2}-\frac{s}{\eta}Z\right)\frac{\Omega_{3}^{2}}{\left(1+r\right)\left(r+2h-1\right)}-2m\omega_{0}=0.
\end{aligned}
\eeq

To the first order in $\frac{\omega\eta}{m}$, Eq. (\ref{TBS2}) becomes
\begin{eqnarray*}
0 & \approx & \frac{1}{m}\left[-2\frac{\omega\eta}{m}sm{\cal E}+\left(Z^{2}-\frac{s}{\eta}Z\right)\frac{\Omega_{3}^{2}}{\left(1+r\right)\left(r+2h-1\right)}\right]\omega_{0}-2\omega_{0}^{2},
\end{eqnarray*}
which implies either 
\beq\label{omega0K}
\omega_{0}=0\ {\mathrm{or}}\ \omega_{0}=-\frac{\omega\eta}{m}s{\cal E}+\frac{K}{m},
\eeq
where
\[
K\equiv\frac{1}{2}\left(Z^{2}-\frac{s}{\eta}Z\right)\frac{\Omega_{3}^{2}}{\left(1+r\right)\left(r+2h-1\right)}.
\]


\renewcommand{\theequation}{B.\arabic{equation}}

\setcounter{equation}{0}

{\section{Appendix: Hard-wall constraint \label{ApB} }

Since it is impossible to implement the hard-wall condition ($F\left(\rho_{0}\right)=0$) exactly for a harmonic oscillator, as stated e.g. in Ref. \cite{ValentimBakke}, hereafter we express the validity of our solution with Heun polynomials in terms of the physical parameters required to obtain the desired statistical significance \cite{StatBook}. The statistical significance quantifies by how much the probability of our wave function lies outside the hard-wall. This should be negligible for a physically meaningful solution: 
\[
\int^{\infty}_{\rho_{0}}d\rho\left|R\left(\rho\right)\right|^{2}\approx0.
\]
We can express $R\left(\rho\right)$ from Eq. (\ref{ChangeVariables}) as 
\[
R\left(\rho\right) = e^{-k\rho^{2}} e^{-q\rho} \rho^h F\left(\rho\right) = e^{\frac{-\left(\rho+\frac{q}{2k}\right)^{2}}{2\sigma^{2}}} e^{\frac{q^{2}}{4k}} \rho^h F\left(\rho\right),
\]
where the standard deviation is given by (see e.g. Ref. \cite{StatBook}):
\[
\sigma\equiv\frac{1}{\sqrt{2k}},
\]
and $F\left(\rho\right)$ is a Heun polynomial. The Gaussian exponential
is the dominant term when $\rho\gg-\frac{q}{2k}$. 
Moreover, the hard-wall is located at $\rho_0 \equiv \frac{1}{\omega \eta}$; its precise position is set by the values of the parameters $\omega$ and $\eta$. Therefore, the probability density will be negligibly small for $\rho > \rho_0$ if $\rho_0$ is sufficiently larger than the expectation value ($-\frac{q}{2k}$). This will depend on the value of $k$ (or equivalently the standard deviation).
For instance, if the hard-wall is $6\sigma$ distant from the expectation value, the probability of finding the particle outside the hard-wall is less than $2\times10^{-7}\%$. Accordingly, we demand
\[
\rho_{0}+\frac{q}{2k}\ge j\sigma=\frac{j}{\sqrt{2k}},\;j\in\mathbb{R},
\]
where the coverage factor $j$ should correspond to the desired significance level: the larger the $j$, the better the confinement. Hereafter, we search for a relation between the factor of confinement $j$ and the parameters of our physical system. After substituting the physical parameters into the previous inequality, we obtain:
\begin{equation}\label{hardwallconstr}
\frac{\left|\left(\omega_{0}+2s{\cal E}\frac{\omega\eta}{m}\right)m^{2}\omega_{0}\right|^{\frac{1}{4}}}{jm}\ge\frac{\omega\eta}{m}\left(1-\frac{\left\{ \omega^{2}\left[-s\eta-2\left(\frac{1}{2}+L\right)\right]-s\frac{m\omega_{0}}{\eta}-\omega{\cal E}\right\} \Omega_{3}}{2j\left|\left(\omega_{0}+2s{\cal E}\frac{\omega\eta}{m}\right)m^{2}\omega_{0}\right|^{\frac{3}{4}}}\right).
\end{equation}

Now we restrict our discussion to $n_{0}=1$ with the approximation $\frac{\omega\eta}{m}\ll1$. If we
 neglect terms of order $O\left(\left(\frac{\omega\eta}{m}\right)^{2}\right)$
and higher, then the previous inequality reads
\[
\frac{1}{j}\frac{K^{2}}{m^{4}}\ge\frac{\omega\eta}{m}\left(\frac{1}{m^{3}}\left|K\right|^{\frac{3}{2}}+s\frac{K}{\eta m^{2}}\frac{\Omega_{3}}{m}\frac{1}{2j}\right).
\]

This condition is immediately satisfied for $s=1$, which leads to $K=0$. The most restrictive
case is the one for $s=-1$, in which $K>0$. Hence
\[
\frac{\Omega_{3}}{\omega}\ge\frac{\left[1+\frac{1}{\eta}\left(\frac{1}{2}+L\right)\right]}{\left[1+\frac{3}{\eta}\left(L+\frac{1}{2}\right)+\frac{2}{\eta^{2}}\left(L+\frac{1}{2}\right)^{2}\right]}\eta^{2}\left(2j\sqrt{\frac{\left[1+\frac{3}{\eta}\left(L+\frac{1}{2}\right)+\frac{2}{\eta^{2}}\left(L+\frac{1}{2}\right)^{2}\right]}{2\left[1+\frac{1}{\eta}\left(\frac{1}{2}+L\right)\right]}}-1\right).
\]
This is understood as a condition on the ratio $\frac{\Omega_{3}}{\omega}$
for the validity of our approximated solution with respect to the hard-wall. This
ratio is constrained by the inequality above according to the values
of the parameters $L$ and $\eta$. The string parameter $\eta$ is
in the interval $(0,1]$. We will specify the above condition in both
limits of the interval and check the conditions for its validity. 

For $\eta\approx1$:
\[
\frac{\Omega_{3}}{\omega}\gtrsim j\sqrt{\frac{2+2\left(\frac{1}{2}+L\right)}{\left[1+3\left(L+\frac{1}{2}\right)+2\left(L+\frac{1}{2}\right)^{2}\right]}},
\]
so either 
\[
\frac{\Omega_{3}}{\omega}\gtrsim 0,\ {\rm for}\,L\rightarrow\infty
\]
or
\[
\frac{\Omega_{3}}{\omega}\gtrsim j,\ {\rm for}\,L=0.
\]
Thus in both situations, the condition can be satisfied by appropriate values of the physical parameters.

On the other hand, in the limit $\eta\rightarrow0$, we find
\[
\frac{\Omega_{3}}{\omega}\gtrsim j\sqrt{\frac{\eta^{5}}{\left(L+\frac{1}{2}\right)}},
\]
so that
\[
\frac{\Omega_{3}}{\omega}\gtrsim 0,\ {\rm for}\,L\rightarrow\infty
\]
or
\[
\frac{\Omega_{3}}{\omega}\gtrsim j\sqrt{2\eta^{5}},\ {\rm for}\,L=0.
\]
We can state that if the condition $\frac{\Omega_{3}}{\omega}\gtrsim j$
is satisfied in these two limits, then our solution will satisfy the
hard-wall constraint.}

\end{document}